\documentclass[amssymb]{article}

\usepackage{graphicx}

\begin{document}

\begin{center}

{\bf \Large Discrete Laplacian Growth: Linear Stability vs Fractal Formation}\\

\vspace{5mm}

{\Large Igor Loutsenko, Oksana Yermolayeva}\\

\vspace{3mm}

ICTP, Trieste, 34014, Italy\\

\vspace{5mm}

e-mail: loutsenk@maths.ox.ac.uk, oyermola@ictp.it

\vspace{5mm}

Abstract

\end{center}

\begin{quote}

We introduce stochastic Discrete Laplacian Growth and consider its
deterministic continuous version. These are reminiscent respectively
to well-known Diffusion Limited Aggregation and Hele-Shaw free
boundary problem for the interface propagation. We study correlation
between stability of deterministic free-boundary problem and
macroscopic fractal growth in the corresponding discrete problem. It turns out
that fractal growth in the discrete problem is not influenced by stability
of its deterministic version. Using this
fact one can easily provide a qualitative analytic description of
the Discrete Laplacian Growth.

\end{quote}

\begin{section}{Introduction}

The Laplacian Growth  or zero surface tension Hele-Shaw free-boundary
problem ( see e.g. \cite{B}, \cite{VE}) describes time evolution of a
domain $\Omega=\Omega(t)$ in the complex plane with $z=x+iy$, when the boundary
$\partial\Omega$ of the domain is driven by the gradient of a scalar
field $\phi=\phi(x,y,t)$ (see Fig. \ref{continuous}). The field is
harmonic, except for several singular points ("sources" and "sinks"), in
the exterior or the interior of $\Omega$. The corresponding
Hele-Shaw problems are referred as exterior and interior
respectively. The field $\phi$ vanishes at the domain boundary
("interface") $\phi(\partial\Omega)=0$. Well possedness and
stability of the problem depends strongly on types of singularities.

For instance, the exterior "sink"-driven Hele-Shaw problem  is a
problem of expanding a simply-connected domain $\Omega$ whose
boundary is driven by the field harmonic in exterior of $\Omega$. It is
ill-posed and linearly unstable almost for any initial conditions.
On the contrary, the exterior problem for contracting $\Omega$ (which is
a "source"-driven time reversal of the above problem) is linearly
stable and well posed.

A discrete, stochastic version of the exterior Hele-Shaw problem
(often called Diffusion Limited Aggregation or DLA,
\cite{GL},\cite{H},\cite{WS}) describes formation of a cluster of
particles on two dimensional (e.g. square) lattice. No more than one
particle can occupy a lattice cite. The particles stick one by one
to the cluster. The probability for a particle to occupy a given
(next to the cluster) cite is proportional to the value of a lattice
harmonic field at that cite. The field has logarithmic "sink"-type
singularity at infinity. It vanishes on the cluster and is updated
after each cluster increment.

The cluster in such a discrete problem is a fractal and one may
think that instability and ill-possedness of the deterministic
continuous version of such a discrete model (i.e. exterior Hele-Shaw
free-boundary flow for expanding domain $\Omega$) is a manifestation
of this fact. In other words, the fractal interface of the discrete
problem may not be approximated by an analytic boundary curve of
its deterministic counterpart.

In this article we consider a "mix" of interior and exterior
problems for simply-connected bounded domain, when the interface is
driven by the scalar field $\phi$ which is harmonic almost everywhere
except for the interface $\partial\Omega$ itself, where $\phi$ vanishes, and
two logarithmic singular points. One of these points is placed at
$z=0$ and other at $z=\infty$. In such systems the growth depends on
a parameter $\lambda$ which is a measure of the "mix" between the exterior
unstable ($\lambda=0$) and the interior stable ($\lambda=1$) processes,
with $\lambda=1/2$ being a "neutral stability" growth.

One may expect that the fractal formations are not visible on
macroscopic scale for discrete version of such models when
$\lambda>1/2$ and when the interface is linearly stable. This turns
out to be true only for extreme stability point $\lambda=1$, while the
macroscopic fractal formations are present for any $\lambda<1$ in
the discrete model.

Note that the Discrete Laplacian  Growth (DLG) introduced here differs from
the Diffusion Limited Aggregation (DLA) by simultaneous consideration of both
exterior and interior processes on the lattice defined in a similar way.
In the case of DLA one considers a lattice
cluster and a complement to it, while in DLG the lattice is divided
into interior domain, discrete boundary and exterior domain.
Although the law of the cluster growth in pure exterior limit
($\lambda=0$) of DLG is locally different from that of DLA,
both models have the same continuous version and belong essentially
to the same class.


In the next section we describe, in details, the continuous version of
the growth processes under consideration and perform its linear
stability analysis.

\end{section}

\begin{section}{Continuous Model}

\begin{figure}
\centering
\includegraphics{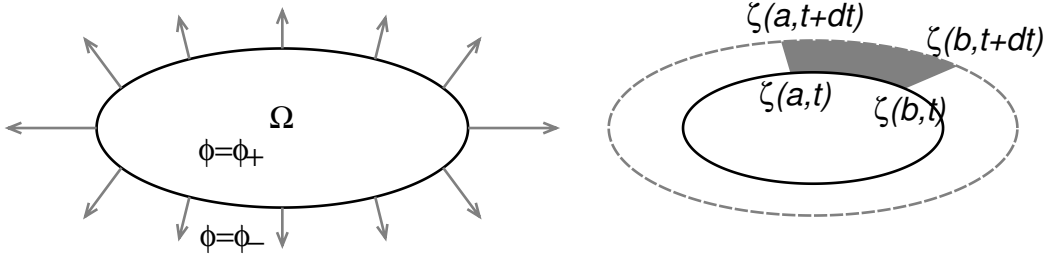}
\caption{Problem setting (left) and the domain area increment along
the boundary segment $a<l<b$ during the time interval $(t,t+dt)$
(right)} \label{continuous}
\end{figure}

Let, for simplicity, $\Omega$ be a simply connected, bounded domain
in the $z=x+iy$ plane with the point $z=0$ inside the domain. We
denote by $\phi_+$ the field $\phi(x,y,t)$ in the interior
and by $\phi_-$ in the exterior of the domain respectively (see Fig.
\ref{continuous}). Field $\phi_\pm$ is harmonic in the
interior/exterior of $\Omega$ except for points $z=0$ and $z=\infty$
\begin{equation}
\Delta \phi_\pm=0, \quad
z\not\in\left\{0,\infty,\partial\Omega\right\}. \label{laplacian}
\end{equation}
It is continuous across the boundary and vanishes on it
\begin{equation}
\phi_\pm(z\in\partial\Omega)=0 \label{boundary}
\end{equation}
The field has logarithmic singularities at $z=0$ and $z=\infty$
$$
\phi_+\to \frac{A_+}{2\pi}\log|z|, \quad z\to 0, \quad
$$
$$
\phi_-\to\frac{A_-}{2\pi}\log|z|, \quad z\to \infty
$$
where $A_\pm$ are constants.

Consider the situation when the boundary dynamics is governed by
gradients of $\phi_\pm$
$$
v_n=\alpha_+\frac{\partial\phi_+}{\partial
n}+\alpha_-\frac{\partial\phi_-}{\partial n},
$$
where $\alpha_\pm$ are constants, $v_n$ denotes the normal velocity
of the boundary
$$
v_n={\rm Re}(n d\bar z/dt), \quad \bar z=x-iy, \quad
z\in\partial\Omega,
$$
$n=n_x+in_y, |n|=1$ stands for the exterior normal to
$\partial\Omega$ and $\partial/\partial n$ denotes the normal
derivative at the boundary.

Rescaling the harmonic fields $\phi_\pm$ and the time variable $t$
$$
\phi_+\to-\frac{A_+}{\lambda}\phi_+, \quad
\phi_-\to\frac{A_-}{1-\lambda}\phi_-, \quad t\to
\frac{\lambda}{\alpha_+A_+}t, \quad
$$
where
$$
\lambda=\frac{A_+\alpha_+}{A_-\alpha_-+A_+\alpha_+},
$$
we rewrite the above dynamic law for the boundary velocity as
\begin{equation}
v_n=\frac{\partial\phi_-}{\partial n}-\frac{\partial\phi_+}{\partial
n} \label{velocity},
\end{equation}
and field asymptotic now depend on a "stability" parameter $\lambda$
as
\begin{equation}
\phi_+\to \frac{-\lambda}{2\pi}\log|z|, \quad z\to 0, \quad
\phi_-\to\frac{1-\lambda}{2\pi}\log|z|, \quad z\to \infty.
\label{sources}
\end{equation}

Equations
(\ref{laplacian},\ref{boundary},\ref{velocity},\ref{sources})
together with the initial condition $\Omega(t=0)$ constitute a
free-boundary, initial value problem for $\Omega(t)$.

Note, that the case $\lambda=0$ corresponds to the exterior Hele-Shaw
problem (with $\phi_+=0$) while $\lambda=1$ refers to the interior Hele-Shaw
problem (with $\phi_-=0$, respectively).

We are interested in discrete stochastic processes of lattice
cluster growth, corresponding to the deterministic problem
(\ref{laplacian}-\ref{sources}).

Consider first the case
of cluster growing from a single particle at origin that
corresponds to the circular solution of deterministic model. It is easy
to see that such solution is an expanding circle of radius
$r(t)=\sqrt{t/\pi}$ and $\phi_\pm=\phi^{(0)}_\pm $, where
$$
\phi_+^{(0)}=\frac{-\lambda}{2\pi}(\log|z|-\frac{1}{2}\log\frac{t}{\pi}),
\quad |z|<r(t), \quad
$$
$$
\phi^{(0)}_-=\frac{1-\lambda}{2\pi}(\log|z|-\frac{1}{2}\log\frac{t}{\pi}),
\quad |z|>r(t)
$$
Let us now study linear stability of this solution, considering
small perturbations of the circle $|z|=r(t)$ by the $k$-periodic
function
$$
|z(\theta,t)|=r(t)+\epsilon r^{(k)}(t)\sin(k\theta)
$$
where $\theta=(0,2\pi)$ is a polar angle on the $z$-plane
($z=|z|\exp(i\theta)$) and $k$ is a positive integer. The fields
$\phi_\pm$ satisfying conditions (\ref{laplacian}), (\ref{sources})
are of the form
$$
\phi_\pm=\phi_\pm^{(0)}+\epsilon\phi^{(k)}(t)|z/r(t)|^{\pm
k}\sin(k\theta).
$$
Substituting this and previous equations to (\ref{boundary}) and
(\ref{velocity}) in the first order of $\epsilon$-series we get
\begin{equation}
\frac{dr^{(k)}}{dt}=\frac{1-2\lambda}{2t}kr^{(k)} \label{analysis}
\end{equation}
Since $k>0$, the perturbations grow in time when $\lambda<1/2$ and
vanish in time when $\lambda>1/2$, with $\lambda=1/2$ being the "neutral
stability" point.

In this article we consider continuous systems labelled by stability
parameter $\lambda$ in the range $0\le\lambda\le
1$. The $\lambda=0$ case (exterior Hele-Shaw problem
for expanding interior domain) is unstable and
ill-posed problem, whose discrete version manifests fractal growth.
On the other extreme of the interval, at $\lambda=1$, we have stable and
well-posed interior expansion problem.

Now it is natural to ask the question: Whether the stability of interior
continuous problems (i.e. those at $\lambda>1/2$) may depress the macroscopic (i.e. visible
in continuous limit) fractal growth of boundaries
of corresponding discrete models?

To study this question, we are to introduce discrete analogue of the above problem
(\ref{laplacian}-\ref{sources}).

\end{section}

\begin{section}{Discrete Model}

One may think of the free boundary problem
(\ref{laplacian}-\ref{sources}) as that of dynamics of an ideal
conducting contour $\partial\Omega$ in two-dimensional electric
field. The field is created by a Coulomb charge of value $-\lambda$
at $z=0$ and unit charge distributed with linear density
$\sigma(l,t)$ along the contour. The contour is "an ideal
conductor", and this means that the potential (at fixed $t$) is
constant along $\partial\Omega$. Here $l$ stands for natural
parameter along the contour $\left\{\partial\Omega: z=\zeta(l,t),
|d\zeta(l)|=dl\right\}$ and $\oint_{\partial\Omega}\sigma(l,t)dl=1$.
The harmonic field $\phi(x,y,t)$ is (modulo coordinate independent
function of $t$) the electrostatic potential created by the above
charges
$$
\phi=\frac{-\lambda}{2\pi}\log|z|+\frac{1}{2\pi}\oint_{\partial\Omega}\sigma(l,t)\log|z-\zeta(l,t)|dl
$$
The gradient of the potential is the electric vector field, that has
jump of magnitude $\sigma(l)$ across the boundary. Therefore from
(\ref{velocity}) it follows that the normal velocity of the contour
equals its linear density
$$
v_n(l,t)=\sigma(l,t)
$$
This and the previous equations reduce the free-boundary problem on the
plane to a dynamical problem on the contour only. It is easy to see that
the density $\sigma$ is non-negative, and the interior domain
$\Omega$ expands in time.

The domain area increment $dP_{(a,b)}$ along the interface segment
$a<l<b$ during the time $dt$ (see Fig. \ref{continuous})
\begin{equation}
\frac{dP_{(a,b)}}{d
t}=\int_{a}^{b}v_n(l)dl=\int_{a}^{b}\sigma(l)dl=q_{(a,b)}
\label{probability}
\end{equation}
is proportional to the electric charge $q_{(a,b)}$ concentrated on
this segment. This fact suggests to consider the boundary charge as
the cluster increment probability in discrete analogue of the above
model.

\begin{figure}[htp]
\includegraphics{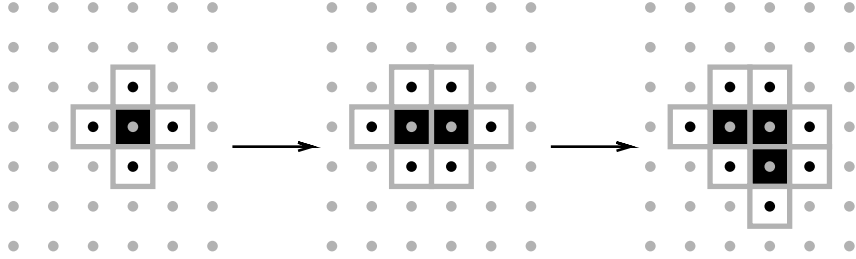}
\caption{An example of the cluster growth. Black squares
("particles") belong to the cluster $\Omega$, white squares with
gray sides form the cluster boundary $\partial\Omega$. Black dots
denote charged (boundary) cites and gray dots are uncharged lattice
cites.}\label{explaincluster}
\end{figure}

Let us now take a square lattice, paint elementary square cells in
white color and label centers of squares by a pair of integers
$(m,n)$ that are discrete coordinates along the $x$ and $y$
directions. Consider the following stochastic process: At the first
step, the $(0,0)$ square is painted in black. At the next step we
paint in black one of four cites that are next-neighbors of this
smallest cluster e.t.c. (see Figure \ref{explaincluster}). This
procedure is continued by coloring, at each step, any randomly
chosen white next-neighbors of the cluster (i.e. adding randomly a
"particle" to the cluster) with the probability described below.

We use the same notation $\Omega$ for the cluster as we used for the
continuous domain. The cluster boundary, which consists of all white
next-neighbors of the cluster, is denoted, by analogy with
continuous case, by $\partial\Omega$.

At each step of the process we can define a lattice function
$\phi_{m,n}$ which satisfies the difference equation
\begin{equation}
\phi_{n-1,m}+\phi_{n+1,m}+\phi_{n,m-1}+\phi_{n,m+1}-4\phi_{n,m}=-\lambda\delta_{n,0}\delta_{m,0},
\quad (n,m)\not\in\partial\Omega \label{lattice_laplacian}
\end{equation}
everywhere except for the boundary, where it vanishes
\begin{equation}
\phi_{n,m}=0, \quad (n,m)\in\partial\Omega. \label{lattice_boundary}
\end{equation}
Its asymptotics is like follows
\begin{equation}
\phi_{n,m}\to\frac{1-\lambda}{4\pi}\log(m^2+n^2), \quad
m^2+n^2\to\infty \label{lattice_sources}
\end{equation}
The left hand side of (\ref{lattice_laplacian}) is a lattice Laplace
operator, and $\delta$ on the right hand side stands for Kronecker
delta symbol. Equations
(\ref{lattice_laplacian}-\ref{lattice_sources}) are lattice analogues
of (\ref{laplacian}, \ref{boundary}, \ref{sources}).

As in the continuous case, the potential $\phi_{n,m}$ can be
expressed in terms of charges, that are now placed at the boundary
cites (see Fig. \ref{explaincluster}).

Introducing the lattice Coulomb potential (or the Green function)
$G_{n,m}$, a such that
$$
G_{n-1,m}+G_{n+1,m}+G_{n,m-1}+G_{n,m+1}-4G_{n,m}=\delta_{n,0}\delta_{m,0},
\quad -\infty<n,m<\infty
$$
and
$$
G_{n,m}\to\frac{1}{4\pi}\log(m^2+n^2), \quad m^2+n^2\to\infty
$$
one expresses $\phi_{n,m}$ in terms of $q_{i,j},
(i,j)\in\partial\Omega$ as
$$
\phi_{n,m}=\sum_{(i,j)\in\partial\Omega}q_{i,j}G_{n-i,m-j}-\lambda
G_{n,m}+C
$$
where $C$ is a constant.

It's easy to see that the boundary charges are nonnegative $q_{i,j}\ge
0$. From (\ref{lattice_laplacian}), (\ref{lattice_sources}) and by the
definition of the Green function it follows that
\begin{equation}
\sum_{(i,j)\in\partial\Omega}q_{i,j}=1. \label{sum}
\end{equation}
By direct analogy with continuous case (c.f. (\ref{probability})) we
now consider $q_{i,j}$ as a probability of the $(i,j)$ square of the
boundary $\partial\Omega$ to join the cluster $\Omega$.

Therefore, at each step of the growth process we have to solve the
linear algebraic system consisting of equation (\ref{sum}) and
\begin{equation}
\sum_{(i',j')\in\partial\Omega}q_{i',j'}G_{i-i',j-j'}-\lambda
G_{i,j}+C=0, \quad (i,j) \in \partial\Omega \label{potential}
\end{equation}
for unknowns $q_{i,j}, (i,j)\in\partial\Omega$ and $C$, then repaint
one of the boundary squares in the black color with the probability
$q_{i,j}$.

Note, that the lattice function for boundary cites satisfies the following
equation
$$
\phi_{i-1,j}+\phi_{i+1,j}+\phi_{j-1,i}+\phi_{j+1,i}=q_{i,j}, \quad
(i,j)\in\partial\Omega
$$
and the cluster increment rule can be also reformulated in terms of
the potential $\phi$ as follows: Probability of the $(i,j)$ boundary cite to
join the cluster equals to the sum of potentials on the next-neighbors of
that cite.

\end{section}

\begin{section}{Fractal and Continuous Growth}

\begin{figure}[htp]
\centering
\includegraphics{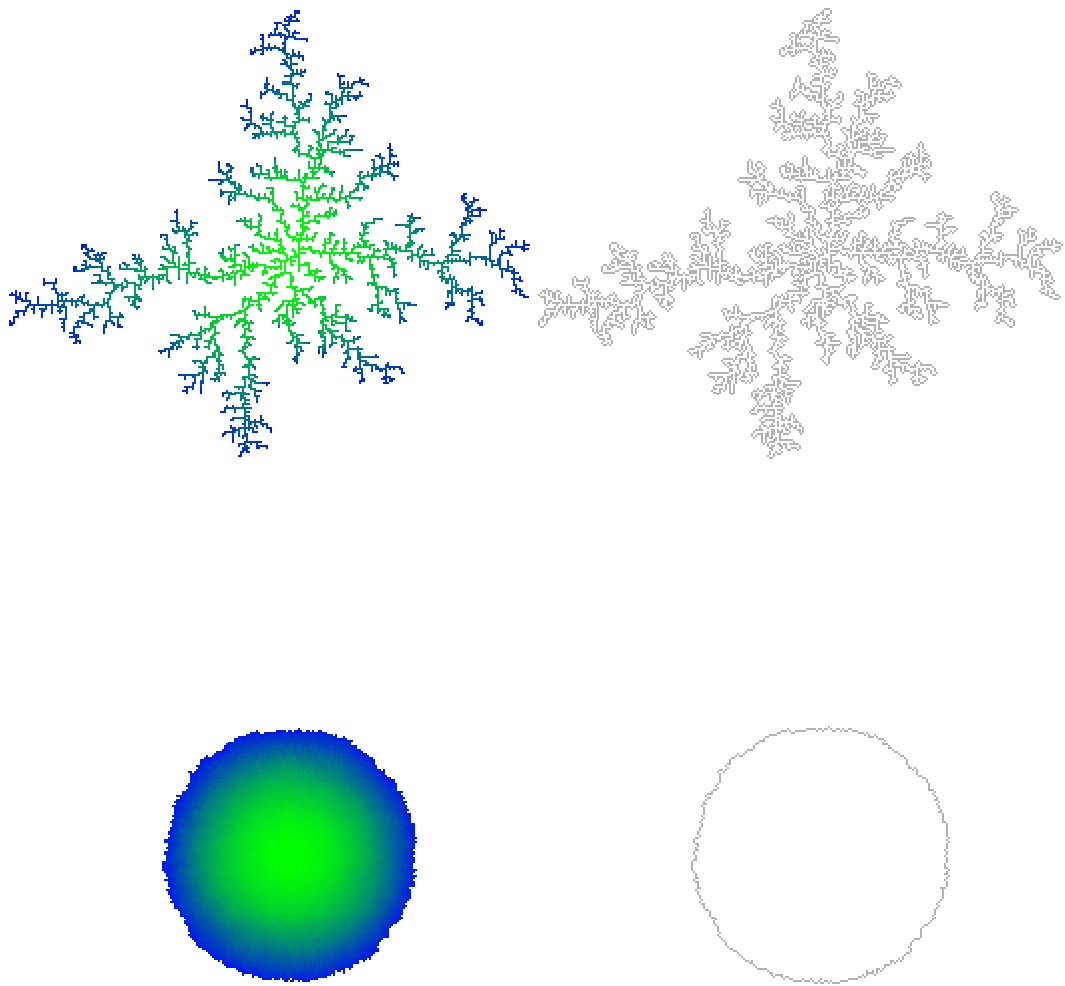}
\caption{Clusters (left) and their boundaries (right) for the pure
exterior  ($\lambda=1$) problem (top) and the pure interior problem
$\lambda=0$ (bottom). To visualize the time dynamics, we prescribe
colors to cluster particles. The color ranges from green to blue,
depending on time (or step) at which the particle joined the cluster
(green for the initial step, blue for the final
step)}\label{extremes}
\end{figure}

\begin{figure}[htp]
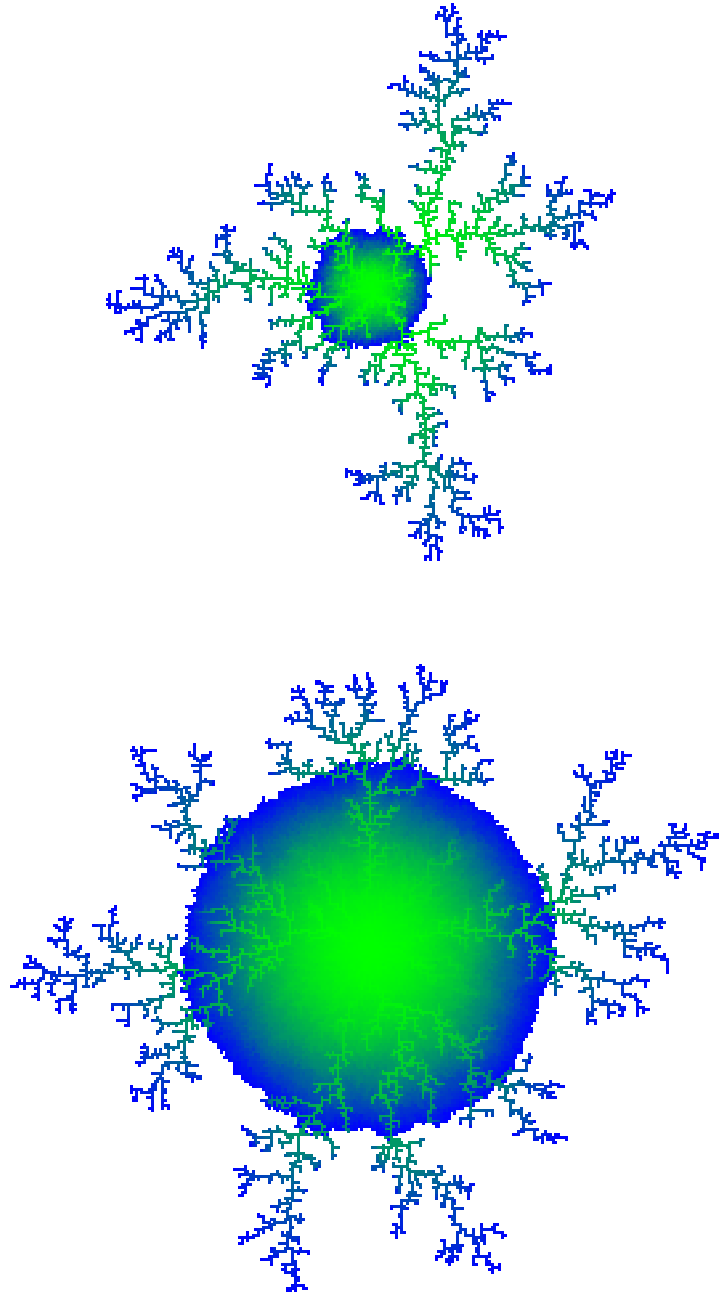

\centering
\includegraphics{fig3_1.ps}
\includegraphics{fig3_2.ps}
\caption{Examples of clusters growing from a single particle for
$\lambda=0.25$ (top) and $\lambda=0.65$ (bottom)}\label{lambdas}
\end{figure}

Since in the case of pure exterior $\lambda=0$ problem the model
under consideration differs from other discretizations of the
Laplacian Growth by local details only, we have to expect the
qualitative behavior to be similar to that of Diffusion Limited
Aggregation. Indeed, for $\lambda=0$ one observes the pure fractal
growth (see Figure \ref{extremes}). Numerical calculations give the
following estimate for Hausdorff dimension of fractal boundary of a
tree-like cluster (see Appendix for details of numerical
calculations)
\begin{equation}
d=1.64\pm0.02 \label{dimension}
\end{equation}
Note that in the $\lambda=0$ case the dimension of the boundary
coincides with that of the cluster.

In the case of the pure interior problem $\lambda=1$, the Discrete
Laplacian Growth shows dynamics, close to that of its deterministic
continuous counterpart: The cluster boundary is not a fractal, and
tends to solution of the continuous problem as its size increases
(see Figure \ref{extremes}). Such a behavior was expected due to
stability of the interior continuous problem for expanding cluster.

Since the continuous problem (\ref{laplacian}-\ref{sources}) is
stable for $\lambda>1/2$ (c.f. (\ref{analysis})) one would expect
similar (close to deterministic) behavior of Discrete Laplacian
Growth for such $\lambda$. This turns out not to be so: Instead, one
observes separation of the growth in two fractal components, one of which is
quasi-regular and the other is macroscopic, i.e. visible in continuous
limit, constituting a finite fraction of the cluster. In the range
$0<\lambda<1$, a big (grown from a single particle) cluster consists
of a quasi-circular center and branches going out of it (see
Figure \ref{lambdas}). The Hausdorff dimension of such cluster
boundary equals the same $d$ as in (\ref{dimension}).

One can interpret such a behavior in the following way: The discrete
growth can be viewed as a competition between the averaged process tending
to a continuous limit, and probabilistic fluctuations. Such
fluctuations drive the system beyond the linear stability region when
$\lambda<1$.

\end{section}

\begin{section}{Qualitative Analysis}

One might try to explain the above fractal properties of growth in
the stability region $1/2<\lambda<1$ by the fact that the cluster
evolution considered so far starts rather with a single particle than
 with quasi-continuous
macroscopic initial conditions.
And indeed, to describe properly the continuous limit we have to
start with a cluster of dimension 2, containing $N\to\infty$
particles. It is also necessary that, after rescaling the lattice
spacing (or cluster linear size) by the factor $N^{-1/2}$, the
initial boundary $\partial\Omega(t=0)$ tends to an analytic curve
as $N\to\infty$. The time, which is the step number in the discrete
model, is rescaled by the factor $1/N$.

Such growth process remains, nevertheless, a
superposition of smooth (analytic) and macroscopic fractal
components in the limit $N\to\infty$: The quasi-analytic part of the
boundary $\partial\Omega(t)$ evolves as in the pure interior problem,
while the quasi-analytic growth is superposed with the fractal growth of
the pure exterior problem. In other words, the "mixed" interior-exterior
problem separates into interior and exterior parts in zero-lattice
spacing limit.

It is easy to find the rate of growth of quasi-analytic to fractal
components provided the separation always takes place for
$0<\lambda<1$: The tree-like fractal part mainly grows on the tips
of the "branches", since the exterior field $\phi_-$ is screened out
in fjords between the branches, while the interior field $\phi_+$ is
screened out inside the branches. Therefore, in zero-lattice spacing
limit, the interior (quasi-analytic) part of the boundary evolves
like there were only one logarithmic singularity
$$
\phi_+\to\frac{- \lambda}{2\pi}\log|z|, \quad z\to 0, \quad {\rm and}
\quad \phi_-=0,
$$
while the fractal part of the curve grows as if
$$
\phi_-\to\frac{1-\lambda}{2\pi}\log|z|, \quad z\to\infty, \quad {\rm
and} \quad \phi_+=0.
$$
The ratio of growth between the fractal and quasi-analytic parts (in
number of particles per unit of time) is
$$
(1-\lambda)/\lambda.
$$

It is also straightforward to show that the separation always takes
place for $0<\lambda<1$. Let's consider the case when $\lambda$ is
close to 1, since the separation of the fractal component here
implies the separation for a smaller positive $\lambda$. Suppose
again that initial cluster has a big size $N\to\infty$ and its
rescaled boundary is close to an analytic curve, so the number of
the boundary cites is of order $\sqrt{N}$.

Consider now a perturbation (fluctuation) of the boundary. Suppose
that the fluctuation linear size (in lattice spacings) is $l$. For
$\lambda$ close to $1$, the charge of the particles on the
fluctuation tip (i.e. most distant from the quasi-analytic part of
the boundary cites of the fluctuation) is of order
$(1-\lambda)l^{1-D/2}/\sqrt{N}$, where $1\le D\le 2$. For instance,
$D=1$ for one-dimensional, "crack"-like fluctuations of length $l$,
while $D=2$ for a "bump"-like 2-dimensional fluctuations of diameter
$l$. Charge of particles on the quasi-analytic part of the boundary
is of order $1/\sqrt{N}$. A fluctuation tend to grow (rather than
collapse) when the probability for a particle to join the cluster at
the tip of fluctuation exceeds the probability for the
quasi-analytic part of the boundary. Since the probability is
proportional to the charge, from the above discussion it follows
that fluctuations tend to grow when their linear size $l$ exceeds
some critical value which is proportional to
$(1-\lambda)^{-\frac{2}{2-D}}$. This value is asymptotically
independent of cluster size $N$.

Since the critical size of fluctuation is $N$-independent, the
possible number of critical fluctuations along the boundary
is proportional to the
boundary length $\sqrt{N}$. It then follows that starting from the
close to analytic boundary, a big cluster develops critical
fluctuations in the number of steps $f(\lambda)\sqrt{N}$, where
$f(\lambda)$ is some function of $\lambda$ only. This number of
steps corresponds to the time $t=f(\lambda)/\sqrt{N}$ in the scaling
limit.

Therefore, in the $N\to\infty$ scaling limit, the cluster develops
critical fluctuations in zero time and the separation of growth into
the fractal and continuous components takes place immediately if
$0<\lambda<1$.

\end{section}

\begin{section}{Conclusions}

In the present work we have introduced the Discrete Laplacian Growth
(DLG) model and studied correlation between the stability of its
continuous version and the fractal growth of boundary. The
stability, turns out, does not guarantee an absence of the fractal
growth. The growth process separates into superposition of
"continuous" and fractal component. In other words, in the
$N\to\infty$ scaling limit, the discrete growth does not tend to
solution of continuous model even if the later is stable and
well-posed.

There remain a few questions to address:

1) Since, in $N\to\infty$ limit, DLG separates in two independent
processes, one may think about implications of such a separation in
the continuous model. For instance, whether this separation can be traced
(in some, possibly "hidden", form) in continuous model or it is just
a consequence of the discretization?

2) It is also known that both, exterior and interior problems possess
an infinite number of conserved quantities (harmonic moments) in the
scaling $N\to\infty$ limit \cite{R,VE}. Does this imply (in view of the
separation) existence of an infinite number of conserved quantities
in "mixed" models?

3) In this article we presented numerical estimate (\ref{dimension})
for the fractal dimension of DLG. Although DLG differs from the
Diffusion Limited Aggregation (DLA) only locally, their fractal
dimensions do not coincide. This result is not unexpected, since
dimension of DLA is sensitive even to the lattice type (see e.g.
\cite{H}) and can not be considered as a general characteristics
of a class of models. It would
be interesting to understand which quantities are universal, i.e.
independent of local details of a discrete scheme.

\end{section}

\begin{section}{Appendix}

Numerical simulations of the Discrete Laplacian Growth have been
performed by updating the discrete boundary (presented by a set of pairs
of integer coordinates
$\left(i(1),j(1)\right)$,$\left(i(2),j(2)\right)$, $\dots,
\left(i(K),j(K)\right)$, where $K$ is the current boundary length)
with the probability given by solutions of linear system
(\ref{sum},\ref{potential}). In this method an explicit solution of
Laplace equation at each step is not needed, since the coefficients of
 Eq. (\ref{potential}) are defined by the coordinates of
the boundary cites only.

The Green function (or lattice Coulomb potential) $G_{n,m}$ can be
computed in several ways. To speed up the calculations, we have used
the following approximation for the lattice Green function
$$
G_{m,n}=\frac{1}{4\pi}\log(m^2+n^2)-\frac{m^4+n^4-6m^2n^2}{24\pi(m^2+n^2)^3}
$$
when the point $(n,m)$ is in the exterior of the square $-L\le n\le
L, -L\le m\le L$, and numerical solution for $G_{n,m}$ in the
interior of this square (with the function value at the square
boundary given by the above approximation). In this approach, the
Green function is represented numerically as (computed once)
$2L+1\times 2L+1$ table extended by the above equation. This
approximation is very precise: for example, for $L=100$ the maximum
deviation from the exact lattice Green function is of order
$10^{-10}$.

We split solution of the system (\ref{sum},\ref{potential}) in two
parts:
$$
\sum_{k'=1}^KG_{i(k)-i(k'),j(k)-j(k')}q^{\rm ex}_{k'}=1, \quad
\sum_{k'=1}^KG_{i(k)-i(k'),j(k)-j(k')}q^{\rm in}_{k'}=G_{i(k),j(k)}
$$
$$
q_k=-C q^{\rm ex}_k+\lambda q_k^{\rm in}, \quad C=-\frac{1-\lambda
\sum_{k=1}^K q_k^{\rm in}}{\sum_{k=1}^K q_k^{\rm ex}},
$$
where $q_k$ stands for $q_{i(k),j(k)}$. The $K\times K$ matrix
$G_{i(k)-i(k'),j(k)-j(k')}$ is symmetric and positive definite, so
we made use of the rapidly converging conjugate gradient method \cite{KA} for
the numerical solution.

\end{section}

\vspace{5mm}

{\bf \Large Acknowledgement}\\

The authors would like to acknowledge useful information and help
received from Professor V.Kravtsov.

\end{document}